\documentclass[intlimits,twoside,a4paper]{article}

\usepackage{graphicx}
\usepackage{amsmath}
\usepackage[T2A]{fontenc}
\usepackage[cp1251]{inputenc}
\usepackage{cmpj}
\usepackage{amssymb}

\issue{2011}{14}{1}{13602}

\doinumber{10.5488/CMP.14.13602}
\title[ Polar phonon states and their dispersive spectra in supperlattices]%
{Polar optical phonon states and their dispersive spectra of a
wurtzite nitride superlattice with
complex bases: transfer-matrix method%
\thanks{The author would like
to acknowledge the valuable guidance and discussion of Prof. J. Shi
of Peking University. This work was supported by the National
Natural Science Foundation of China for Youth under Grant
No. 60906042.%
}}
\author[L.~Zhang]{L.~Zhang\thanks{E-mail:
zhangli-gz@263.net}}
\address{ Department of Mechanism and Electronics, Guangzhou Panyu Polytechnic,
511483 Guangzhou, People's Republic of China}
\authorcopyright{L.~Zhang}
 \date{Received September 3, 2010, in final form October 31, 2010}

\begin{document}

\maketitle

\begin{abstract}
Based on the dielectric continuum model and transfer-matrix
method, the completing polar optical phonon states in a wurtzite
GaN-based superlattice (SLs) with arbitrary-layer complex bases
are investigated. It is proved that $2^n$ types of phonon modes
probably exist in a wurtzite nitride SL with $n$-layer complex
bases. The analytical phonon states of these modes and their
dispersive equations in the wurtzite GaN/Al$_{x}$Ga$_{1-x}$N SL
structures are obtained. Numerical calculations on a three-layer
GaN/Al$_{0.15}$Ga$_{0.85}$N/AlN complex bases SL are performed.
Results reveal that there are interface optical (IO) phonon modes
of one type only and four types of quasi-confined (QC) phonon
modes in three-layer GaN/Al$_{0.15}$Ga$_{0.85}$N/AlN complex bases
SLs. The dispersive spectra of phonon modes in complex bases SLs
extend to be a series of frequency bands. The behaviors of QC
modes reducing to IO modes are observed. The present theoretical
scheme and numerical results are quite useful for analyzing the
dispersive spectra of completing phonon modes and their polaronic
effect in wurtzite GaN-based SLs with complex bases.
\pacs 63.22.Np, 68.65.Cd, 71.55.Eq


\end{abstract}

\section{Introduction}
Since the superlattices (SLs) notion was first brought forward
nearly four decades ago~\cite{EsakiL}, the electronics and
opto-electronics properties of semiconductor SLs have attracted
considerable attention due to their fancy miniband effect and a
lot of excellent physical properties~\cite{EsakiL,XiaJB}. Among
them, semiconductor SLs based on group-III nitride GaN, AlN, InN,
and their ternary compounds Al$_x$Ga$ _{1-x}$N, In$_x$Ga$_{1-x}$N
have invoked special interest both in theoretical and
experimental
investigations~\cite{Darakchieva0,OrtegaMFA,Gil,Nakamura1,Yuls}.
This is mainly ascribed to the following evident characteristics: the
nitride materials with widely and adjustable direct-band gap,
strong atomic bonding and high carrier mobility make them
attractive materials as a basis for the creation of reliable
high-temperature and high-frequency electronic equipment and
short-wavelength optoelectronics
devices~\cite{Gil,Nakamura1,Yuls}.

In order to further improve the performance of microelectronics
devices based on SL structures, people brought forward the
complex-bases SL, i.e., $n$ layers per period
($n>2$)~\cite{BoudoutiEE,RouhaniaD,EsakiL1}. In contrast to the
binary (two-layer per period) SLs, the complex bases SLs provide
more degrees of freedom for the adjustable structural parameters,
the miniband and minigap widths due to the introduction of
additional layers in each SL period. This is very important for
several important applications, such as infrared photodetectors,
effective-mass filtering, and tuning of the tunneling
current~\cite{YuhP,VasilopoulosP,ChoiKK}. The previous research
on the complex bases SL is mainly focused on the electronic
structures and surface polariton properties in GaAs-based
semiconductors~\cite{JJShi01,BoudoutiEH,KucharczykR,SteslickaM,CaiM,MendialduaJ}.
But there are few investigations of the other physical
properties, such as the crystal-lattice dynamical feature and
optoelectronic characteristics in GaN-based SL with complex
bases~\cite{MedeirosSK}. Indeed, crystal-lattice vibrations
(phonon modes) have great effect on the electronic and impurity
states, the excitonic recombination and carrier transport,
especially in low-dimensional quantum systems of polar
semiconductors~\cite{ShiJJ9,XieHJ}. At room and higher
temperatures, the electron-phonon interactions and scattering by
optical phonons play a dominant role effecting various properties of
polar semiconductor quantum heterostructures, including
hot-electron relaxation rates, interband transition rates,
room-temperature exciton lifetimes, and many other optical and
transport properties~\cite{ZhangXB,BalandinAA}. Moreover, since
nitride material usually crystallizes in the anisotropic hexagonal
wurtzite structure, the crystal dynamical properties of nitride
SLs are more complex due to anisotropy of the crystal structure in
contrast to those of cubic
crystals~\cite{Shi2,Shi3,KomirenkoSM,LiL,ZhangL9,ZhangL0,ZhangL1,ZhangL8}.
Hence, it is necessary and important to investigate the polar
optical phonon modes in wurtzite GaN-based SL with complex bases.

On the basis of the dielectric continuum approximation and
Loudon's uniaxial crystal model~\cite{Loudon}, several authors
have made their great contributions to the study of the polar
optical phonons and their electron-phonon interactions in wurtzite
nitride binary SLs~\cite{MedeirosSK,Gleize0,AnselmoD}. For
example, Gleize and his coworkers~\cite{Gleize0} investigated the
anisotropy effects on polar phonon modes in wurtzite GaN/AlN SLs,
and the quasi-confined (QC) phonon modes and interface-optical
(IO) phonon modes were found in the SL structures. Medeiros,
Anselmo and their cooperators studied the confinement of polar
optical phonons in two-layer periodic GaN/AlN SLs~\cite{AnselmoD}
based on the dielectric continuum model (DCM), and the phonon
dispersive spectra of the systems were displayed in their
discussions. From the viewpoint
 of experimental investigations, Darakchieva et
al.~\cite{DarakchievaV1} studied the phonons in AlN/GaN SLs using a
combination of infrared spectroscopic ellipsometry and Raman
scattering spectroscopy. And the E$_1$(TO), A$_1$(LO) and E$_2$
localized, and E$_1$(LO) and A$_1$(TO) delocalized superlattice
modes were identified. Recently, Davydov et al.~\cite{DavydovVY}
studied the lattice dynamics and Raman spectra of strained hexagonal
wurtzite GaN/AlN and GaN/AlGaN SLs, and thus the IO and the QC phonon
modes were observed. Gleize and coauthors~\cite{Gleize2} probed
the confined phonons in hexagonal GaN-AlN artificial SL structures
based on non-resonant Raman scattering, and the QC optical phonon
modes were observed. Gleize's group~\cite{Gleize3} also measured
the angular dispersion of polar IO and QC phonons in a hexagonal
nitride SL, and the experimental data were in good agreement with
the results of a previous calculation based on a DCM. Dutta et
al.~\cite{DuttaM} reported Raman scattering results in binary SLs
of GaN/(In)AlN, which also proves the availability of DCM for the description of polar optical phonon
modes in wurtzite nitride heterostructures. Therefore, the DCM and
Loudon's uniaxial crystal model are reliable and will be adopted
to the investigation of the optical phonon modes in wurtzite nitride SL
with complex bases in the present paper.

The main accomplishments and significance of this work lie in the
following three points. (i) Via solving the Laplace equation of
phonon potentials in wurtzite SLs with $n$-layer complex bases, it
is confirmed that $2^n$ types of polar phonon modes, i.e. the
propagating (PR) mode, the IO mode, the ($2^n-2$) QC modes
probably coexist in wurtzite complex-base SL structures. (ii) By
using the transfer-matrix method~\cite{BoudoutiEE,RouhaniaD}, the
polar optical phonon states and corresponding dispersive equations
for the $2^n$ types of polar phonon modes in wurtzite nitride
complex-bases SLs are given within the framework of the DCM and
the Loudon's uniaxial lattice model. (iii) As an example, a
wurtzite GaN/Al$_x$Ga$_{1-x}$N/Al$_y$Ga$_{1-y}$N SL with
three-layer complex bases is chosen to check the theoretical
scheme described in the present paper. The numerical result
reveals that there are five types of phonon modes in the wurtzite
GaN/Al$_x$Ga$_{1-x}$N/Al$_y$Ga$_{1-y}$N three-layer SLs chosen
here. The dispersive spectra behave as a series of frequency
bands, which are narrower than those in wurtzite two-layer
GaN-based SLs~\cite{MedeirosSK,Gleize0,AnselmoD}. The present
theoretical scheme and numerical results are important and useful
for further experimental and theoretical investigations of the
polar phonon modes behaviors and their effect on optoelectronic
properties in wurtzite nitride SLs with complex bases. The rest
sections of the paper are organized as follows: the phonon states
and their dispersive equations of a wurtzite GaN-based SL with
$n$-layer complex bases are deduced in section~2 based on the
transfer-matrix method; the numerical results for the dispersion
relation, the electrostatic potential functions of the phonon
modes on a wurtzite GaN/Al$_x$Ga$_{1-x}$N/Al$_y$Ga$_{1-y}$N SL
with three-layer complex bases are performed and discussed in
section~3; Finally, we summarized the main\linebreak results and discussed
the significance of the theory described in the current work in
the last section.

\section{Theory}

Let us consider a wurtzite
GaN/Al$_{x_1}$Ga$_{1-{x_1}}$N/Al$_{x_2}$Ga$_{1-{x_2}}$N/
$\ldots$/Al$_{x_{n-1}}$Ga$_{1-{x_{n-1}}}$N quasi-2-dimensional
(Q2D) complex bases SL with $n$-layer per period (as shown in
figure~1). The $z$ axis is taken along the [0001] direction
($c$-axis) of the wurtzite crystals, and the origin of coordinate
is chosen at the left hand side of some a GaN layer. The widths of GaN
layers and the Al$_{x_i}$Ga$_{1-{x_i}}$N layers are $d_i$
($i=0,\,1,\,2,\,3,\ldots,\,n-1$), respectively. Thus the period of
the complex bases SL is $L=\sum_{i=0}^{n-1} d_i$\,. For
convenience, we label the GaN layer with origin of coordinate and
its right Al$_{x_i}$Ga$_{1-{x_i}}$N ($i=1,\,2,\ldots,\,n-1$)
layers as the unit cell 0. The periodic unit cells of the SL on
the right hand side of the unit cell 0 are labeled successively as unit
cell $1,\,2,\,3,\ldots$\,, while those on the left hand side of unit
cell 0 are labeled as unit cell $-1,-2,-3,\ldots$\,. Due to the
anisotropy of wurtzite crystals, the polar phonon frequencies and
the dielectric function becomes direction dependent. If one
denotes the direction of $c$-axis as $z$ and its perpendicular
direction as $t$, the dielectric tensor can be written as
\begin{eqnarray}
\epsilon _i(\omega ) &=&\left(
\begin{array}{lll}
\epsilon _{t,i}(\omega ) & 0 & 0 \\
0 & \epsilon _{t,i}(\omega ) & 0 \\
0 & 0 & \epsilon _{z,i}(\omega )
\end{array}
\right),  \label{1}
\end{eqnarray}
where
\begin{equation}
\epsilon _{t,i}(\omega )=\epsilon _{t,i}^\infty \frac{\omega
^2-\omega _{t,{\rm L},i}^2}{\omega ^2-\omega _{t,{\rm
T},i}^2}\,,\qquad
\epsilon _{z,i}(\omega )=\epsilon _{z,i}^\infty \frac{\omega
^2-\omega _{z,{\rm L},i}^2}{\omega ^2-\omega _{z,{\rm T},i}^2}\,,
\qquad(i=0,\,1,\,2,\ldots,\,n-1). \label{5}
\end{equation}
Here $\omega _{z,{\rm L}}$, $\omega _{z,{\rm T}}$, $\omega
_{t,{\rm L}}$ and $\omega _{t,{\rm T}}$
are the zone center characteristic frequencies of A$_1$(LO), A$_1$(TO), E$_1$%
(LO), and E$_1$(TO) modes, respectively. The subscripts
$i=0,\,1,\,2,\ldots,\,n-1$ denote the $i$th-layer
Al$_{x_i}$Ga$_{1-{x_i}}$N materials in any unit cell $N$,
respectively.

 \begin{figure}[htb]
\centerline{\includegraphics[width=0.9\textwidth]{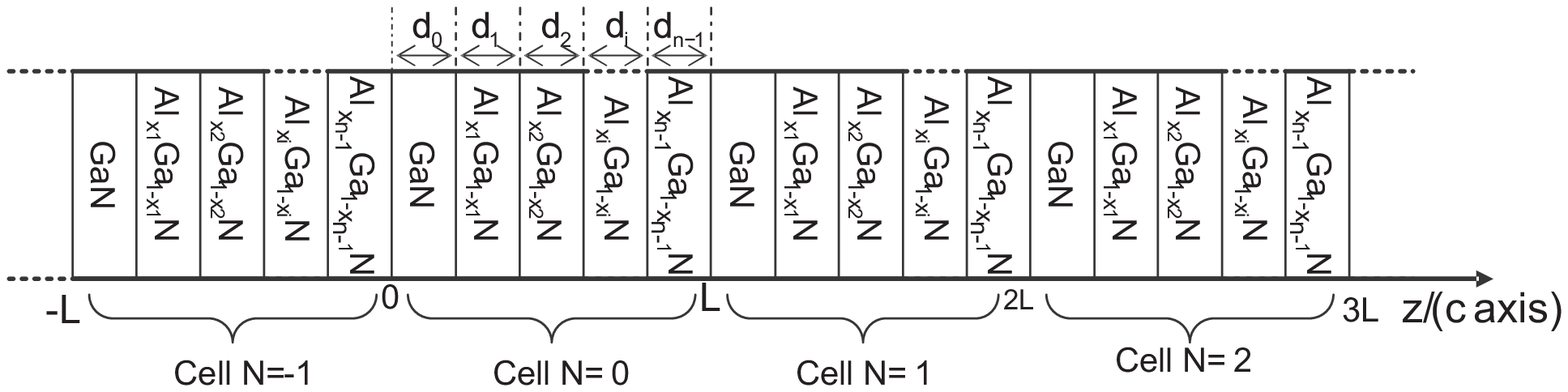}}
\caption{Schematic drawing of a GaN-based SL with $n$-layer
complex bases. Here the $z$-axis is taken along the [0001]
direction ($c$-axis).} \label{fig-smp1}
\end{figure}

In the case of free oscillations (the charge density $\rho _0(%
\mathbf{r})=0$) and using the Maxwell equations, the electric
displacement vector $\mathbf{D}$ satisfies the following equation
under the cylindrical coordinates, i.e.
\begin{equation}
\mathbf{\nabla }\cdot \mathbf{D}=-\varepsilon _0\left\{ \epsilon
_t(\omega )\left[ \frac 1\rho \frac \partial {\partial \rho }(\rho
\frac \partial
{\partial \rho })+\frac 1{\rho ^2}\frac{\partial ^2}{\partial \varphi ^2}%
\right] +\epsilon _z(\omega )\frac{\partial ^2}{\partial z^2}\right\} \Phi (%
\mathbf{r})=0.  \label{25}
\end{equation}
It is obvious that equation~(\ref{25}) is a Laplace equation, and
its solution $\Phi (\mathbf{r})$ is the electrostatic potential of
polar optical phonon modes. In fact, via solving the
equation~(\ref{25}) in the wurtzite SLs, one can get the complete
polar optical phonon modes. In order to solve equation~(\ref{25})
conveniently, we define two functions as
\begin{eqnarray}
\zeta _i(\omega ) &=&\mathrm{sign}\left[ \epsilon _{z,i}(\omega
)\epsilon
_{t,i}(\omega )\right] ,  \nonumber \\
\gamma _i(\omega ) &=&\sqrt{\epsilon _{t,i}(\omega )/\epsilon _{z,i}(\omega )}%
,\qquad(i=0,\,1,\,2,\ldots,\,n-1),  \nonumber \\
\gamma _i(\omega ) &=&\left\{
\begin{array}{lll}
\quad\sqrt{\left| \epsilon _{t,i}(\omega )/\epsilon _{z,i}(\omega
)\right| }, &  &
\zeta _i(\omega )\geqslant 0, \\
\pm{\rm i}\sqrt{\left| \epsilon _{t,i}(\omega )/\epsilon
_{z,i}(\omega )\right| }, &  & \zeta _i(\omega )<0.
\end{array}
\right.   \label{24}
\end{eqnarray}
In terms of the theories of two-order differential equations, we
know that, when $\zeta _i(\omega )<0$, the electrostatic
potentials of phonon modes in the $i$th-layer material correspond
to the oscillating waves. On the contrary, they correspond to the
decaying ones when $\zeta _i(\omega
)>0$~\cite{Shi3,KomirenkoSM,LiL}. Taking into account the
oscillating or decaying waves in $n$-layers of some unit cell in
the complex bases SL structures, it can be confirmed that $2^n$
types of polar phonon modes are likely to exist in wurtzite GaN-based SLs
with $n$-layer complex bases. They are the IO phonon mode ($\zeta
_{i}(\omega )>0$), the  PR phonon mode ($\zeta _{i}(\omega
)<0$)~\cite{Shi2,Gleize0}, and the ($2^n-2$) types of the QC
phonon modes (at least one $\zeta _i(\omega )>0$, and one $\zeta
_j(\omega )<0$, $i\neq j$)~\cite{Gleize0,DavydovVY,Gleize3} in
wurtzite SLs, respectively. These QC modes are distinguished and
labeled as follows. In some unit cell of the complex bases SL,
if the function $\zeta _{i}(\omega )$ takes negative values in
$j={\rm u,v,w,\ldots}$ ($j<n$) layers, then the QC mode is
expressed by QC$^{\rm u,v,w,\ldots}$\,. For example, in a
three-layer complex bases SL, the QC modes are denoted as QC$^{\rm
i-ii}$ if the functions $\zeta _{i}(\omega )$ ($i=1,\,2$) are
negative, and $\zeta _{3}(\omega )>0$.

Next, let us discuss the phonon states and dispersive equations of
all the phonon modes in wurtzite nitride SLs with $n$-layer complex
bases in a uniform method. Considering the translational invariance
in ${\bf \rho}$-plane (the perpendicular plane to $z$ axis) of the
SL quantum structures, the phonon electrostatic potentials in the
$N$-th unit cell of the wurtzite GaN-based SLs with complex bases
can be written as follows,
\begin{equation}
\Phi _N(\mathbf{r})=\Phi _N(\mathbf{\rho },z)=\mathrm{e}^{\mathrm{i}\mathbf{k%
}_t\cdot \mathbf{\rho }}\phi _N(z),  \label{30}
\end{equation}
and
\begin{eqnarray}
\phi _N(z)&=&\left\{
\begin{array}{lll}
\phi _{N,0}(z), &  & NL<z \leqslant NL+d_0\,, \\
\ldots &  & \ldots \\
\phi _{N,i}(z), &  & NL+\sum_{i=0}^{i-1}d_i<z \leqslant NL+\sum_{i=0}^{i}d_i\,, \\
\ldots &  & \ldots \\
\phi _{N,n-1}(z), &  & NL+\sum_{i=0}^{n-2}d_i<z\leqslant (N+1)L
\end{array}
\right.   \label{38} \\
&=&\mathrm{e}^{\mathrm{i}qNL}\times \left\{
\begin{array}{l}
A_0\exp [\gamma _0k_t(z-NL)]+B_0\exp [-\gamma _0k_t(z-NL)], \\
\ldots \\
A_i\exp [\gamma _ik_t(z-NL-\sum_{i=0}^{i-1}d_i)] \\
+B_i\exp [-\gamma _ik_t(z-NL-\sum_{i=0}^{i-1}d_i)], \\
\ldots \\
A_{n-1}\exp [\gamma _{n-1}k_t(z-NL-\sum_{i=0}^{n-2}d_i)] \\
+B_{n-1}\exp [-\gamma _{n-1}k_t(z-NL-\sum_{i=0}^{n-2}d_i)],
\end{array}
\right.   \nonumber
\end{eqnarray}
where ${\bf k}_t$ is a wave-vector of phonon modes in the
$\rho$-plane. And $q$ is a wave vector that will arise in the
dispersion relation for the collective excitations in SLs. In
fact, the phase factor $\mathrm{e}^{\mathrm{i}qNL}$ is introduced
to fulfill the Bloch's theorem:~\cite{StreightSR,TsuruokaT}
\begin{eqnarray}
\phi (z) &=&\mathrm{e}^{\mathrm{i}qz}U_q(z),  \nonumber \\
U_q(z+NL) &=&U_q(z),\quad(N=0,\pm1,\pm2,\ldots).  \label{39}
\end{eqnarray}
Via equation~(\ref{24}), it is seen that $\gamma _i(\omega
)=\sqrt{\left| \epsilon _{t,i}(\omega )/\epsilon _{z,i}(\omega
)\right|}$ as $\zeta _i(\omega )>0$. The potential $\phi
_{N,i}(z)={\cal A}_i\exp [\gamma _ik_tz] +{\cal B}_i\exp [-\gamma
_ik_t z]$ in the $i$-layer of unit cell $N$ behaves as a decaying
wave based on equation~(\ref{38}) in this situation. On the
contrary, $\gamma _i(\omega )=\pm{\rm i}\sqrt{\left| \epsilon
_{t,i}(\omega )/\epsilon _{z,i}(\omega )\right|}$ as $\zeta
_i(\omega )<0$. Under this situation, based on
equation~(\ref{38}), it is found that the potential function $\phi
_{N,i}(z)={\cal A}_i\exp [\,{\rm i}\left| \gamma _i\right|k_tz]
+{\cal B}_i\exp [-{\rm i}\left|\gamma _i\right| k_t z]$ in the
$i$-layer of unit cell $N$, which just behaves as an oscillating
wave. This illustrates that, whether $\zeta _i(\omega )$ is
positive or negative, the equation~(\ref{38}) gives the complete
phonon electrostatic potentials of the phonon modes including the
IO modes, the PR modes and the QC modes in wurtzite complex bases
SLs.

The phonon potential functions and the normal components of
electric displacement should be continuous at each heterostructure
interface. These are referred to as continuity  boundary
conditions (BCs). Using these continuity BCs, one can obtain a
series of BC Eqs. (\ref{41}) in GaN-based SLs with complex bases,
i.e.,
\begin{eqnarray}
\phi _{N,0}(NL) &=&\phi _{N-1,n-1}(NL)\,,  \nonumber \\
\epsilon _{z,0}\frac{\mathrm{d}\phi
_{N,0}(z)}{\mathrm{d}z}\Big|_{z=NL} &=&\epsilon
_{z,n-1}\frac{\mathrm{d}\phi
_{N-1,n-1}(z)}{\mathrm{d}z}\Big|_{z=NL}\,,
\nonumber \\
\ldots&\ldots&\ldots  \nonumber \\
\phi _{N,i}(NL+\sum_{i=0}^{i-1}d_i) &=&\phi
_{N-1,i+1}(NL+\sum_{i=0}^{i-1}d_i)\,,  \nonumber \\
\epsilon _{z,i}\frac{\mathrm{d}\phi _{N,i}(z)}{\mathrm{d}z}%
\Big|_{z=NL+\sum_{i=0}^{i-1}d_i} &=&\epsilon
_{z,i+1}\frac{\mathrm{d}\phi
_{N-1,i+1}(z)}{\mathrm{d}z}\Big|_{z=NL+\sum_{i=0}^{i-1}d_i}\,,  \nonumber \\
\ldots&\ldots&\ldots  \nonumber \\
\phi _{N,n-2}(NL+\sum_{i=0}^{n-2}d_i) &=&\phi
_{N,n-1}(NL+\sum_{i=0}^{n-2}d_i)\,,  \nonumber \\
\epsilon _{z,n-2}\frac{\mathrm{d}\phi _{N,n-2}(z)}{\mathrm{d}z}%
\Big|_{z=NL+\sum_{i=0}^{n-2}d_i} &=&\epsilon
_{z,n-1}\frac{\mathrm{d}\phi
_{N,n-1}(z)}{\mathrm{d}z}\Big|_{z=NL+\sum_{i=0}^{n-2}d_i}\,.
\label{41}
\end{eqnarray}
Substituting equation~(\ref{38}) into equations~(\ref{41}), and
defining two $2\times2$ matrixes equations~(\ref{82}), (\ref{85})
and a vector (\ref{87}), the equations~(\ref{41}) can be expressed
in concise forms. The matrixes and the vector are defined as follows:
\begin{equation}
\mathbf{M}_i=\left(
\begin{array}{cc}
\exp (\gamma _ik_td_i) & \exp (-\gamma _ik_td_i) \\
\epsilon _{z,i}\exp (\gamma _ik_td_i) & -\epsilon _{z,i}\exp (\gamma
_ik_td_i)
\end{array}
\right),   \label{82}
\end{equation}
\begin{equation}
\mathbf{N}_i=\left(
\begin{array}{cc}
1 & 1 \\
\epsilon _{z,i}\gamma _i & -\epsilon _{z,i}\gamma _i
\end{array}
\right),   \label{85}
\end{equation}
and
\begin{equation}
\mathbf{C}_i=\left(
\begin{array}{l}
A_i \\
B_i
\end{array}
\right).   \label{87}
\end{equation}
Using the above definitions (\ref{82}) to (\ref{87}), the
equations~(\ref{41}) can be expressed compactly by the following
matrix equations~(\ref{91}), i.e.,
\begin{eqnarray}
\exp (-{\rm i}qL)\mathbf{M}_{n-1}\mathbf{C}_{n-1}
&=&\mathbf{N}_0\mathbf{C}_0\,,
\nonumber \\
\ldots&\ldots&\ldots,  \nonumber \\
\mathbf{M}_i\mathbf{C}_i &=&\mathbf{N}_{i+1}\mathbf{C}_{i+1}\,,
\quad (i=0,\,1,\ldots,\,n-2)  \label{91} \\
\ldots&\ldots&\ldots  \nonumber \\
\mathbf{M}_{n-2}\mathbf{C}_{n-2}
&=&\mathbf{N}_{n-1}\mathbf{C}_{n-1}\,. \nonumber
\end{eqnarray}
Via solving inverse matrix on equation~(\ref{91}), it is easy to
get the relation:
\begin{equation}
\exp (-{\rm i}qL)\mathbf{M}_{n-1}\mathbf{C}_{n-1}=\prod_{j=0}^{n-1}\mathbf{D}_j%
\mathbf{C}_{n-1}\,,  \label{93}
\end{equation}
where
\begin{equation}
\mathbf{D}_j=\mathbf{M}_j^{-1}\mathbf{N}_{j+1}\,.  \label{95}
\end{equation}
The condition of the linear and homogeneous equations~(\ref{93})
having nonzero solutions for ${\bf C}_{n-1}$ will result in
$2\times2$ coefficient determinant equal to 0, i.e.,
\begin{equation}
{\rm Det}[\exp (-{\rm
i}qL)\mathbf{M}_{n-1}-\prod_{j=0}^{n-1}\mathbf{D}_j]=0. \label{97}
\end{equation}
This equation~(\ref{97}) is just the dispersive equation of phonon
modes in wurtzite nitride SLs with complex bases. Substituting the
dielectric functions (\ref{5}) and the structural parameters $d_i$
into the dispersive equation~(\ref{97}), the dispersive
frequencies of phonon modes in the complex based nitride SLs can
be worked out as a function of the wave-number $k_t$\,. It is
worth to mention that, as the materials become isotropic (i.e.
$\epsilon_t=\epsilon_z$), the dispersive equation~(\ref{97}) of
anisotropic SL crystals will reduce to those of isotropic
SLs~\cite{StroscioMA0}. In the situation, there are two types of
phonon modes, i.e., the IO and the longitudinal-optical-like modes
exist in the isotropic SL systems~\cite{ShiJJ9,XieHJ,StroscioMA0}.

\section{Numerical results and discussion}

In section~2, we have derived the phonon states and their
dispersive equations of all the polar optical phonon modes in
wurtzite nitride SLs with $n$-layer complex bases. However, the
corresponding analytical formulas are complicated. In order to get
a clear picture for the dispersive behaviors of the types of polar
optical phonon modes in wurtzite complex bases SL systems, an
application of the present theories based on a nitride SL with
GaN/Al$_{0.15}$Ga$_{0.85}$N/AlN three-layer complex bases is
performed, and a numerical calculation of dispersive relationship
is carried out. The material parameters used in our calculations
are listed in table~1~\cite{Shi3,KomirenkoSM,Gleize0}.

\begin{table}[!h]
\tabcolsep 0pt \caption{Zone-center energies (in meV) of polar
optical phonons, dielectric constants of wurtzite GaN,
Al$_{0.15}$Ga$_{0.85}$N  and AlN
materials~\cite{Shi3,KomirenkoSM,Gleize0}.} \vspace*{12pt}
\begin{center}
\def\temptablewidth{0.75\textwidth}
{\rule{\temptablewidth}{1pt}}
\begin{tabular*}{\temptablewidth}{@{\extracolsep{\fill}}cccccc}
& $\omega_{t,{\rm T}}$ & $\omega_{t,{\rm L}}$ & $\omega_{z,{\rm T}}$ &
$\omega_{z,{\rm L}}$ &
$\epsilon_{ \infty}$  \\
      \hline
 {\rm GaN}     & 69.25 & 91.83   &   65.91  & 90.97 & 5.35 \\
 {\rm AlN}     & 83.13 & 113.02   &   75.72  & 110.30 & 4.77 \\
{\rm Al$_{0.15}$Ga$_{0.85}$N}& 71.332 & 90.009   &  67.382  & 93.87 & 5.263  \\
       \end{tabular*}
       {\rule{\temptablewidth}{1pt}}
       \end{center}
       \end{table}

\begin{table}[htb]
\caption{Signs of the functions $\zeta_i(\omega)$ [$i=1$ (GaN), 2
(Al$_{0.15}$Ga$_{0.85}$N) and 3 (AlN)] and the frequency ranges of
the phonon modes in wurtzite three-layer
GaN/Al$_{0.15}$Ga$_{0.85}$N/AlN complex bases SLs.}
\centerline{\includegraphics[width=0.7\textwidth]{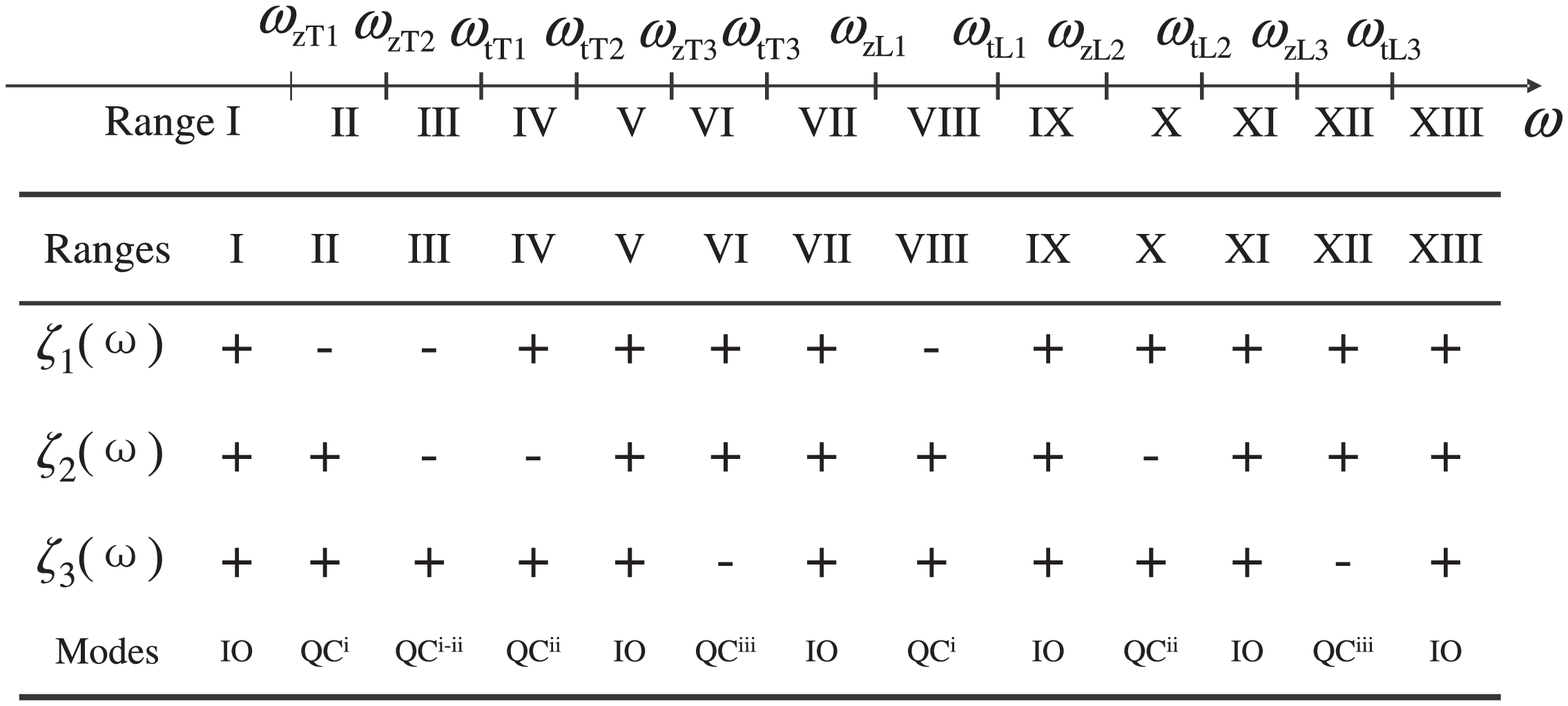}}
 \label{fig-smp0}
\end{table}

Before analyzing the dispersive properties of phonon modes, let us
first discuss the  frequency \nolinebreak ran\-ges of the possible phonon modes in
wurtzite  SL  systems with three-layer
GaN/Al$_{0.15}$Ga$_{0.85}$N/AlN complex bases. In terms of the
material parameters given in table~1, the signs of the functions
$\zeta_i(\omega)$ [$i=1$ (GaN), 2(Al$_{0.15}$Ga$_{0.85}$N),
3(AlN)] and the frequency ranges of the phonon modes in wurtzite
GaN-based SLs with three-layer complex bases are shown in table~2.
From the table, it is observed that the characteristic frequencies
of GaN, Al$_{0.15}$Ga$_{0.85}$N and AlN materials divide the
frequency $\omega$ axis into thirteen subranges. For convenience,
these subranges are labeled as range I, II, $\ldots$, XII, XIII
from left to right. In each frequency subrange, the signs of the
characteristic functions $\zeta_i(\omega)$ [equation~(\ref{5})]
are defined. Based on the properties of different types of phonon
modes, namely behaving as decaying (vibrating) waves in the
complex bases material layers of the SLs, the possible frequency
ranges of these phonon modes in GaN/Al$_{0.15}$Ga$_{0.85}$N/AlN
complex bases SLs can be recognized. It is found that the IO modes
of the system are likely to exist in the subranges I, V, IX, XI
and XIII due to the positive values of $\zeta_{i}(\omega)$
($i=1,\,2,\,3$) in these five subranges. The PR phonon modes do
not appear in the GaN/Al$_{0.15}$Ga$_{0.85}$N/AlN three-layer
complex bases SLs because there are no frequency subranges
satisfying simultaneously the conditions $\zeta_{i}(\omega)<0$
($i=1,\,2,\,3$). It is found that four types of QC phonon modes,
namely, the QC$^{\rm i}$, QC$^{\rm i,ii}$, QC$^{\rm ii}$ and
QC$^{\rm iii}$ phonon modes are likely to appear in the structure.
Here the superscript ``i, ii, iii'' of QC$^{\rm i,ii,iii}$ modes
represent GaN, Al$_{0.15}$Ga$_{0.85}$N and AlN in some unit cell
of the nitride SLs, respectively. In fact, the QC$^{\rm i}$ modes
appear in ranges II and VIII, the QC$^{\rm i,ii}$ mode just exists
in subrange of III, the QC$^{\rm ii}$ modes also appear in two
subranges of IV and X, and the frequency of QC$^{\rm iii}$ modes
falls into the subranges VI and XII. Taking into account the
condition of solution~\cite{Exp}, the dispersive
equation~(\ref{97}) for IO phonon modes has no solution in the
subranges I, VII and XIII. Thus, the IO phonon modes just appear
in the three frequency subranges of V, IX and XI. And the
subranges I, VII and XIII are termed as the forbidden frequency
subranges. Apart from the forbidden frequency subranges I, VII and
XIII of IO modes, the dispersive equations of other QC phonon
modes have solutions in corresponding frequency ranges.
Considering the forbidden subrange VII, the frequency subranges
II, III,$\ldots$, VI are termed as low-frequency range, while the
subranges VIII, IX,$\ldots$, XIII are referred to as
high-frequency range.

 \begin{figure}[htb]
\centerline{\includegraphics[width=0.9\textwidth]{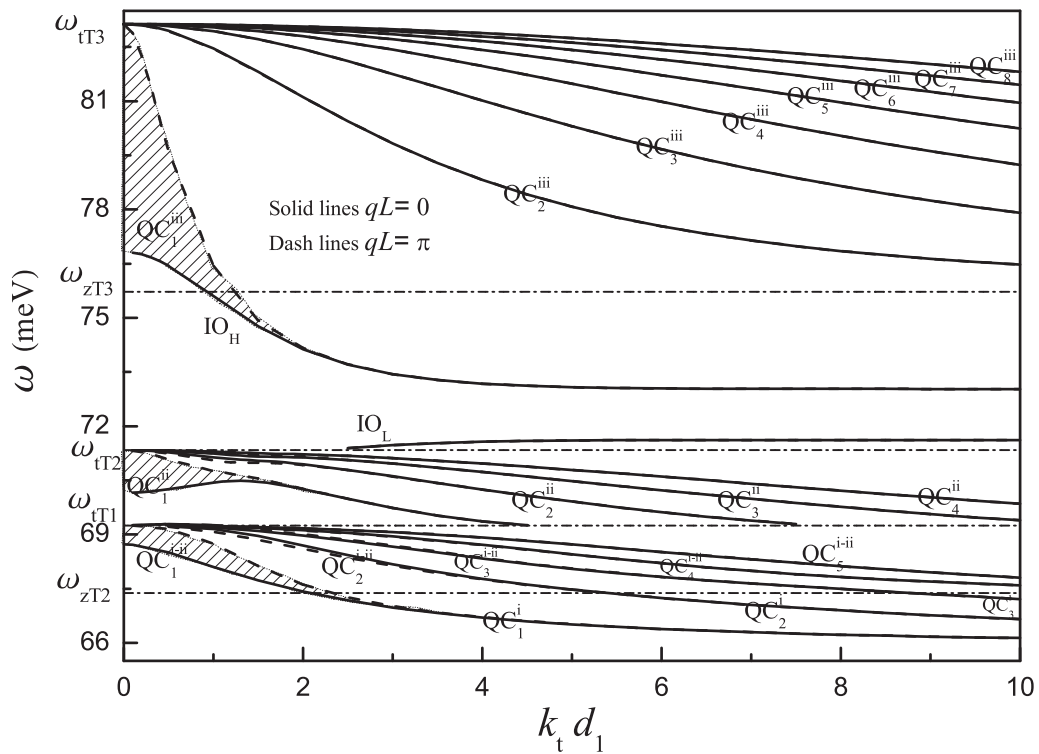}}
\caption{Dispersive spectra of the polar optical phonon modes in
low-frequency as functions of the free wave-number $k_t$ of
$\rho$-plane in a wurtzite three-layer
GaN/Al$_{0.15}$Ga$_{0.85}$N/AlN complex bases SL with
$d_1=d_2=d_3=5$ nm. The solid lines and dashed lines correspond to
the cases of $qL=0$ and $\pi$, respectively.} \label{fig-smp2}
\end{figure}

Figure~2 depicts the dispersive properties of the polar optical
phonon modes in low-frequency range as a function of the free
wave-number $k_t$ of $\rho$-plane in a three-layer
GaN/Al$_{0.15}$Ga$_{0.85}$N/AlN complex bases SL with
$d_1=d_2=d_3=5$ nm. The solid lines and dashed lines correspond to
the cases of phase $qL=0$ and $\pi$, respectively. The areas
between a solid line and an adjacent dashed line are shaded. In fact, these shaded areas are the dispersive frequency
bands of corresponding phonon modes in SLs. From the figure, it is
 clearly seen that four characteristic frequencies $\omega_{z,{\rm
T},2}$\,, $\omega_{t,{\rm T},1}$\,, $\omega_{t,{\rm T},2}$ and
$\omega_{z,{\rm T},3}$ (dashed and dot lines) divide the
low-frequency range into five subranges. In terms of the order of
increasing frequency, the QC$^{\rm i}$, the QC$^{\rm i-ii}$, the
QC$^{\rm ii}$, the IO and the QC$^{\rm iii}$ phonon modes appear
successively in the five frequency subranges, respectively. These
observations are completely consistent with the above analysis in
table~2. Apart from the IO phonon modes, all the other phonon
modes have infinity phonon branches in each frequency subrange.
But only two branches of IO phonon modes exist in the
low-frequency range, and they are labeled by IO$_{\rm L}$ and
IO$_{\rm H}$ shown in the figure. For clarity, no more than six
branches of QC phonon modes are plotted in each frequency subrange
of QC modes here. These QC phonon branches are labeled by QC$^{\rm
i/i-ii/ii/iii}_i$, ($i=1,\,2,\,3,\ldots$), respectively. The
subscript $i$ in symbol QC$^{\rm i/i-ii/ii/iii}_i$ denotes the
order of these QC phonon branches. In figure~1, all the other
dispersive curves are monotonous and decreasing functions of $k_t$
except for the IO$_{\rm L}$ modes and QC$^{\rm ii}_1$ phonon
branch. With the increase of $k_t$\,, the frequency bands
(shaded areas) of phonon modes become narrower and narrower. And
the solid lines and corresponding dashed lines converge to the same
frequency values as $k_t d_1$ approaches 10. The profound physical
reason for this feature lies in the obvious fact that, as the
wave-number is large enough, the wave-length of phonon is quite
short, and the complex bases SLs can be looked at as a series of
independent three-layer GaN/Al$_{0.15}$Ga$_{0.85}$N/AlN QWs. Thus,
the phonon frequency bands of wurtzite SLs reduce to the phonon
branches of wurtzite QWs. For example, the frequency of IO$_{\rm
L}$ modes of the SLs approaches 71.67 meV as $k_t
d_1\rightarrow10$, which is just the same as the IO phonon limited
frequency in GaN/AlN QWs~\cite{Shi2,KomirenkoSM,ZhangL9}.
Similarly, the QC$^{\rm i-ii}$ modes in the SLs will reduce to the
analogous behaviors of PR phonon modes in
GaN/Al$_{0.15}$Ga$_{0.85}$N QWs~\cite{Shi3}. In a certain
frequency subrange of some phonon modes, the higher is the order
$i$ of QC$^{\rm i/i-ii/ii/iii}_i$ ($i=1,\,2,\,3,\ldots$), the
weaker is the dispersion of the phonon modes. This is a general
feature of phonon modes in low-dimensional quantum
structures~\cite{Shi3,KomirenkoSM,Gleize0}. An interesting feature
is that, as the frequencies of QC$^{\rm i-ii}_i$ phonon frequency
bands are lower than the characteristic frequency $\omega_{z,{\rm
T},2}$\,, these QC$^{\rm i-ii}_i$ phonon bands degenerate into the
corresponding QC$^{\rm i}_i$ phonon frequency bands. Also, the QC$^{\rm iii}_1$ degenerating to IO$_{\rm H}$ is clearly observed. But the QC$^{\rm ii}$ phonon modes are cut off at the
frequency $\omega_{t,{\rm T},1}$\,, which means that the number of
the phonon branches decreases with the increase of $k_t$\,. These
are the typical features of phonon modes in anisotropic wurtzite
confined quantum systems~\cite{Shi2,Shi3,KomirenkoSM,Gleize0}. The
extension of frequency bands (shaded areas) for the QC phonon
modes is more distinct relative to those for the IO phonon modes
in the low-frequency range.

 \begin{figure}[htb]
\centerline{\includegraphics[width=0.9\textwidth]{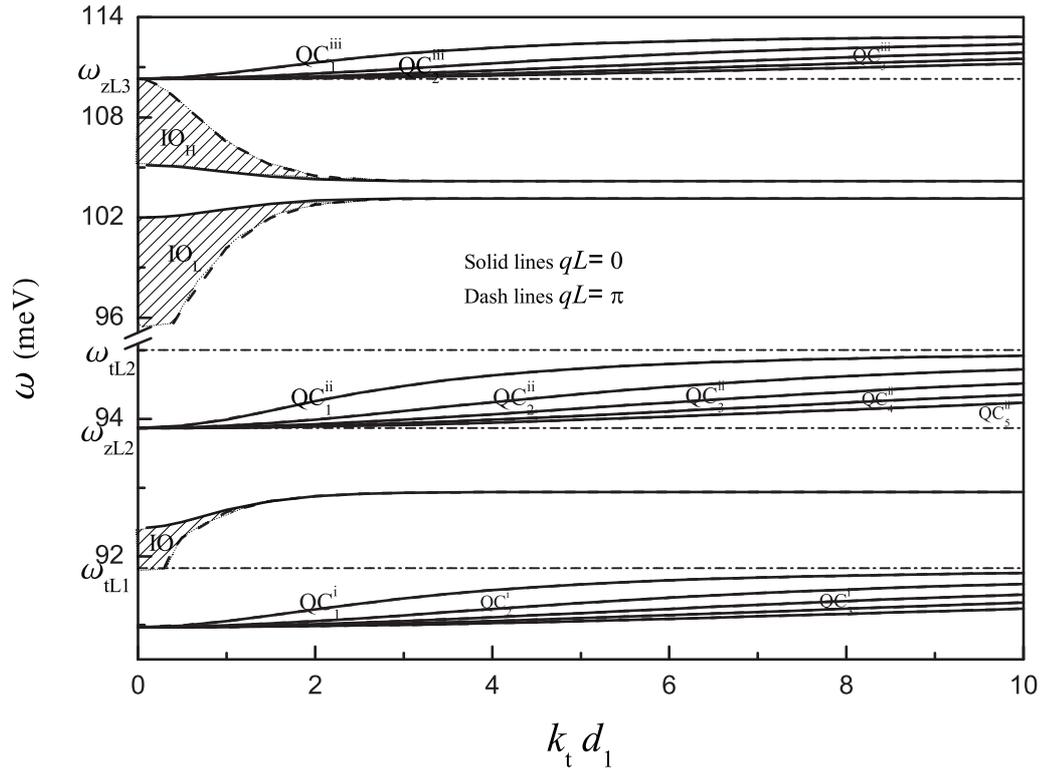}}
\caption{Dispersion frequencies $\omega $ of the polar optical
phonon modes in high-frequency range as functions of $k_t$ in three-layer
GaN/Al$_{0.15}$Ga$_{0.85}$N complex bases SL structures.The
meanings of all curves and symbols in the figure are the same as
those in figure~2.} \label{fig-smp3}
\end{figure}

The dispersive frequencies $\omega$ of the polar optical phonon
modes in high-frequency range as functions of  $k_t$ in a
three-layer GaN/Al$_{0.15}$Ga$_{0.85}$N/AlN complex bases SL are
plotted in figure~3. The meanings of all curves and symbols in the   
figures are the same as those in figure~2. Same as in figure~2,
four characteristic frequencies $\omega_{t,{\rm L},1}$\,,
$\omega_{z,{\rm L},2}$\,, $\omega_{t,{\rm L},2}$ and
$\omega_{z,{\rm L},3}$ (dashed and dotted lines) divide the
high-frequency range into five subranges. The QC$^{\rm i}$ modes
and QC$^{\rm i}$ modes cover the lowest- and the highest-
frequency subranges of the high-frequency range, respectively. The
QC$^{\rm ii}$ phonon modes occupy the middle subrange of
[$\omega_{z,{\rm L},2}\,, \omega_{t,{\rm L},2}$]. These are the
same as in figure~2. But the IO phonon modes of high-frequency
range appear in two subranges, which is quite different from those
in figure~2. Except for the IO$_{\rm H}$\,, the other dispersive
curves are monotonous and increased functions of $k_t$\,. As $k_t
d_1\rightarrow 10$, the IO phonon modes in the two subranges
converge to three certain frequency values, which are the same
value of IO phonon modes in GaN/Al$_{0.15}$Ga$_{0.85}$N/AlN
QWs~\cite{Shi2,KomirenkoSM,ZhangL9}. The physics of this feature
is analyzed above in figure~2. Comparing the frequency bands of
phonon modes in the five subranges, it is found that the frequency
band extensions of the IO modes in contrast to those of the QC
modes in high-frequency range are quite obvious. This is just
opposite to the case of the low-frequency range in figure~2, in
which the frequency band extensions of the QC modes are more
distinct than those of IO phonon modes in low-frequency range. In
fact, the frequency band extension of phonon modes is due to the
periodic crystal structure of SLs, which is quite similar to the
situation of energy band of periodic crystal
structures~\cite{KittelC}. As the wurtzite
GaN/Al$_{x}$Ga$_{1-x}$N/AlN SL reduces to the finite-layer
GaN/Al$_{x}$Ga$_{1-x}$N/AlN QW structures, the QC (IO) phonon
bands in wurtzite SLs will  naturally reduce to the dispersive
curves of QC, PR, half-space (HS) and IO phonon modes in wurtzite
QWs~\cite{Shi2,Shi3,KomirenkoSM,LiL,ZhangL9}.

As stated in the Introduction, the complex bases SLs
possess more adjustable structural parameters, and excellent
miniband and minigap properties. These characteristics utilized in infrared photodetectors, effective-mass filtering,
and tuning of the tunneling
current~\cite{YuhP,VasilopoulosP,ChoiKK}. As the first step to
investigate the polaronic effect on the optic and electronic
properties in GaN-based SL with complex bases, the phonon modes
and their dispersive spectra of the structures are necessary. As a
new and important topic, the geometrical-parameter increases
the polaronic effect of electronic and optoelectronic characteristics in
GaN-based SLs with complex bases and will be studied and reported in
the future. The IO and QC modes in binary GaN/AlN SLs were
observed in experiments~\cite{DavydovVY,Gleize2,Gleize3}. The
experimental observations on phonon modes in GaN-based SLs with
complex bases have been rarely reported by now due to the
difficulty of crystal growth. With the great advance of
semiconductor technology, the phonon modes are anticipated to be found
in the GaN-base SLs with complex bases by the analogous techniques
as those used in~\cite{DavydovVY,Gleize2,Gleize3}. At last, we
should point out that the present theoretical scheme and numerical
results are based on non-retardation limit for simplicity. In
fact, this treatment is widely adopted to be dealt with the crystal
dynamics in GaN-based quantum
structures~\cite{Shi2,Shi3,KomirenkoSM,LiL,Loudon,Gleize0}. The
experimental results of angular dispersion of polar phonons and
Raman scattering have been proven to be in good agreement with the
calculations based on the non-retarded DCM~\cite{Gleize2,Gleize3}.
The recent calculation of polaronic binding energy in GaN nanowire
also well agrees  with relative experimental values~\cite{ZhangL7}.
This shows the reliability and meaningfulness of non-retardation treatment
for the phonon modes in GaN quantum structures. Of course, a
general case of retarded modes can be considered within the framework of
transverse magnetic (TM) polarization~\cite{Loudon}.

\section{Conclusions}

In the present paper, we have analyzed and discussed the
completing polar optical phonon modes in a wurtzite GaN-based SL
with arbitrary-layer complex bases using the transfer-matrix
method. Via solving the Laplace equations in wurtzite crystal, it
is proved that $2^n$ types of polar phonon modes including the PR,
IO, QC$^{\rm i/ii/iii,\ldots}$ phonon modes are likely to exist in
wurtzite nitride SLs with $n$-layer complex bases. The analytical
phonon states of these phonon modes are obtained by means of the
DCM and Loudon's uniaxial crystal model. Based on appropriate BCs,
the concise dispersive equations of these phonon modes in wurtzite
complex bases SLs are derived. An application of the present
theories to a wurtzite SL with three layer
GaN/Al$_{0.15}$Ga$_{0.85}$N/AlN complex bases is performed.
Numerical calculations on dispersive properties are carried out.
Our results reveal that there are five types of phonon modes, i.e.,
one type of IO mode and four types of QC modes coexisting in the
three-layer GaN/Al$_{0.15}$Ga$_{0.85}$N/AlN complex bases SLs. The
dispersive spectra of these phonon modes in SLs extend to be a
series of frequency bands. It is also pointed out that, if the SL
structures degenerate into corresponding finite wurtzite QWs, the
QC and IO phonon bands in wurtzite SLs will naturally reduce to
the PR, IO, HS and QC phonon modes in wurtzite
QWs~\cite{Shi2,Shi3,KomirenkoSM,LiL,ZhangL9}. The present
theoretical and numerical results are important for further
analysis and discussion of the dispersive spectra of phonon modes
and their polaronic effect on the optical and electronic
properties in wurtzite GaN/Al$_x$Ga$_{1-x}$N SLs. We hope that the
present work will stimulate further theoretical and experimental
investigations of phonon properties, as well as their effect on
the physical properties in wurtzite GaN-based SL systems.

%
%

\ukrainianpart

\title[ Polar phonon states and their dispersive spectra in supperlattices]%
{Стани полярних оптичних фононів та їхні  дисперсійні спектри вюрцит нітридної супергратки  зі складними базисами: метод трансфер-матриці%
}
\author[L.~Zhang]{Л.~Жанг}
\address{Політехніка Гуанчжоу Панью, Гуанчжоу, Народна Республіка Китай}

 \date{Отримано 3 вересня 2010р., в остаточному вигляді --- 31 жовтня 2010р.}

\makeukrtitle

\begin{abstract}
\tolerance=3000%
На основі діелектричної неперервної моделі і методу трансфер-матриці, досліджено доповнені стани полярних оптичних фононів у вюрцит  GaN-базованій супергратці (СГ) зі складними базисами.
Доведено, що  $2^n$ типи фононних мод існують у вюрцит нітридній СГ з  $n$-шаровими складними базисами. Отримано аналітичні фононні стани цих мод та їхні дисперсійні рівняння  у структурах вюрциту  GaN/Al$_{x}$Ga$_{1-x}$N СГ. Здійснено числові розрахунки на тришарових
GaN/Al$_{0.15}$Ga$_{0.85}$N/AlN складних базисах СГ. Результати показують, що є інтерфейсні оптичні  (ІО) фононні моди тільки одного типу і чотири типи квазіобмежених (КО) фононних мод у тришарових   GaN/Al$_{0.15}$Ga$_{0.85}$N/AlN складних базисах суперграток. Дисперсійні спектри фононних мод в складних базисах суперграток розширюються до низки частотних зон. Спостерігається, що поведінка КО мод редукується до IО мод. Ця теоретична схема і числові результати є досить корисними для аналізу дисперсійних спектрів доповнених фононних мод і їхніх поляронних впливів у вюрцит GaN-базованих супергратках із складними базисами.
\keywords оптичні фононні моди, дисперсійні спектри, супергратка вюрциту, метод трансфер-матриці


\end{abstract}

\begin{thebibliography}{99}
\bibitem{Exp} The three dielectric functions $\epsilon
_i(\omega )$ ($i=1,\,2,\,3$) (\ref{5}) for GaN,
Al$_{0.15}$Ga$_{0.85}$N and AlN materials take positive or
negative values simultaneously in the frequency subranges of I,
VII and XIII. For example, in the subrange I, all the functions
$\epsilon _{j,i}(\omega )$ ($j=z,t;\,i=1,\,2,\,3$) are positive.
Under this situation, the dispersive equation~(\ref{97}) has no
solution in the subranges I, VII and XIII. This means that the
frequency subranges I, VII and XIII are not the frequency ranges
of for IO phonon modes. Thus the IO modes just appear in the
subranges of V, IX and XI.
\end{thebibliography}
\end{document}